# Crystalline Electric Field Effects in CeMIn$_5$: Superconductivity and the Influence of Kondo Spin Fluctuations

A. D. Christianson[1,2], E. D. Bauer[2], J. M. Lawrence[1], P. S. Riseborough[3], N. O. Moreno[2], P. G. Pagliuso[2], J. L. Sarrao[2], J. D. Thompson[2], E. A. Goremychkin[4], F. R. Trouw[2], M. P. Hehlen[2], and R. J. McQueeney[2,5]

[1]University of California, Irvine, California 92697
[2]Los Alamos National Laboratory, Los Alamos, New Mexico 87545
[3]Temple University, Philadelphia, Pennsylvania 19122
[4]Argonne National Laboratory, Argonne, Illinois 60439
[5]Iowa State University, Ames, Iowa 50011

We have measured the crystalline electric field (CEF) excitations of the CeMIn$_5$ (M = Co, Rh, Ir) series of heavy fermion superconductors by means of inelastic neutron scattering. Fits to a CEF model reproduce the inelastic neutron scattering spectra and the high temperature magnetic susceptibility. The CEF parameters, energy level splittings, and wavefunctions are tabulated for each member of the CeMIn$_5$ series and compared to each other as well as to the results of previous measurements. Our results indicate that the CEF level splitting in all three materials is similar, and can be thought of as being derived from the cubic parent compound CeIn$_3$ in which an excited state quartet at ~12 meV is split into two doublets by the lower symmetry of the tetragonal environment of the CeMIn$_5$ materials. In each case, the CEF excitations are observed as broad lines in the inelastic neutron scattering spectrum. We attribute this broadening to Kondo hybridization of the localized f moments with the conduction electrons. The evolution of the superconducting transition temperatures in the different members of CeMIn$_5$ can then be understood as a direct consequence of the strength of this hybridization. Due to the importance of Kondo spin fluctuations in these materials, we also present calculations within the non-crossing approximation (NCA) to the Anderson impurity model including the effect of CEF level splitting for the inelastic neutron scattering spectra and the magnetic susceptibility.

## I. Introduction

The discovery of the family of CeMIn$_5$ (M = Co, Rh, Ir) heavy fermion superconductors has sparked great interest[1,2,3,4,5]. This is in large part due to the small number of heavy fermion superconductors available for study, the unusually high T$_c$ of 2.3 K observed for CeCoIn$_5$,[3] and the presence of superconductivity and magnetism in the same crystal structure. The substitution of different transition metals (Co, Rh, or Ir) affects the nearest neighbor environment of the Ce$^{3+}$ ion both through small changes in the position of the neighboring ions and through differing hybridization of the Ce f-electron with the conduction electrons. This allows for comparison among the different members of the family where many of the complications normally encountered in the study of heavy fermion materials can be taken to be approximately the same.

All members of the CeMIn$_5$ family crystallize in the tetragonal HoCoGa$_5$ crystal structure (space group P4/*mmm*) and can be viewed structurally as being composed of alternating layers of CeIn$_3$ and MIn$_2$. At ambient pressure, CeCoIn$_5$ and CeIrIn$_5$ are superconducting at 2.3[3] and 0.4 K[2] respectively. On the other hand, CeRhIn$_5$ undergoes an antiferromagnetic transition at 3.8 K and upon application of pressure becomes superconducting at 2.1 K and 16 kbar coinciding with a suppression of the Néel order.[1] The origin of superconductivity in these materials remains poorly understood. However, there is substantial evidence of the unconventional nature of the superconductivity including power-law behavior in the low temperature specific heat and thermal conductivity[6,7] and the spin lattice relaxation rate[8,9,10].

A number of proposals have been made with respect to the origin of heavy fermion superconductivity in the CeMIn$_5$ series. A prominent view is that the CeMIn$_5$ materials are in close proximity to a quantum critical point (QCP)[11,12,13,14,15] where the relevant tuning parameter is the f-conduction electron hybridization. Near the QCP, the effect of substitution of one transition metal for another is to change the hybridization in an analogous manner to the effect of applied pressure on CeIn$_3$.[16] The strong magnetic fluctuations present near a QCP are then implicated as the analog to phonons in conventional BCS superconductivity. Furthermore, the suggestion has been made based upon both experimental and theoretical grounds that crystalline electric field (CEF) effects are important for the heavy fermion ground states found in these materials. It has been argued that the symmetry of the ground state CEF doublet in these materials may be directly relevant to the f-conduction electron hybridization and in some cases may produce spin fluctuations which are more favorable to the formation of the superconducting condensate.[17] In another proposal, CEF splitting affects the competition between spin and orbital fluctuations that, in turn, controls the ground state configuration.[18]

A number of attempts have been made based upon bulk measurements to elucidate the CEF splittings in the CeMIn$_5$ series.[17,19,20,21,22] In each case there is significant discrepancies in the reported results. To clarify the role of CEF excitations and to resolve the discrepancies found in previous experiments we have performed inelastic neutron scattering (INS) experiments to directly probe the CEF excitations in the CeMIn$_5$ series.

For Ce$^{3+}$ in a tetragonal environment, as is the case for theCeMIn$_5$ family, the CEF Hamiltonian can be written as

$$H_{CEF} = B_2^0 O_2^0 + B_4^0 O_4^0 + B_4^4 O_4^4 \tag{1}$$

where the $B_l^m$ are the CEF parameters and the $O_l^m$ are the Stevens operator equivalents. Diagonalization of this Hamiltonian yields the following wavefunctions.[23]

$$\Gamma_7^1 = \alpha\left|\pm\tfrac{5}{2}\right\rangle - \beta\left|\mp\tfrac{3}{2}\right\rangle \quad \Gamma_7^2 = \beta\left|\pm\tfrac{5}{2}\right\rangle + \alpha\left|\mp\tfrac{3}{2}\right\rangle \quad \Gamma_6 = \left|\pm\tfrac{1}{2}\right\rangle \tag{2}$$

The energy splitting of these levels is determined from the positions of the peaks in the INS spectra, while the mixing parameters (α and β) determined from the amplitude of the peaks.

Our results indicate that the CEF splitting in all three materials is similar and can be thought of as being derived from the cubic parent compound CeIn$_3$ (ref. 24, 25, and 26) in which the $\Gamma_7$ groundstate is retained (in tetragonal symmetry denoted above as $\Gamma_7^1$) and the excited state $\Gamma_8$ quartet at ~12 meV is split into two doublets (the $\Gamma_7^2$ to lower energy and the $\Gamma_6$ to higher energy) by the lower symmetry of the tetragonal environment of the CeMIn$_5$ materials. The amount of mixing of the $J_z$ = 3/2 and 5/2 states in the $\Gamma_7^1$ and $\Gamma_7^2$ states also varies from that of CeIn$_3$. Moreover, the CEF excitations are observed as broad lines in the INS spectrum. We attribute this broadening as the signature of the Kondo hybridization of the localized f moments with the conduction electrons. The evolution of the superconducting transition temperatures in the different members of CeMIn$_5$ can then be understood as a direct consequence of the strength of this hybridization. Due to the importance of Kondo spin fluctuations in these materials, we also present calculations within the non-crossing approximation (NCA) to the Anderson impurity model including the effect of CEF level splitting for the INS spectra and the magnetic susceptibility.

## II. Experimental and Theoretical Details

Large high quality single crystals of CeMIn$_5$ and the nonmagnetic analogues LaMIn$_5$ were obtained using the flux-growth method.[27] In the case of CeCoIn$_5$ and YCoIn$_5$, polycrystalline samples were obtained by heating stochiometric amounts of the constituent elements in an alumina crucible sealed within a quartz tube to 1100 °C and cooling to 900 °C and quenching in liquid nitrogen. The samples were then annealed at 600 °C for 3 weeks. After annealing, the samples were etched in dilute HCl to remove excess free In; magnetic susceptibility measurements indicate the free In content to be less than 2%.[28] Unfortunately, similar attempts at producing LaCoIn$_5$ were unsuccessful. The resulting samples were powdered and placed in a rigid flat plate aluminum sample holder. This sample geometry served to not only minimize the effect of the strong neutron absorption of Rh, Ir, and In but maintained a uniform sample distribution enabling an accurate absorption correction.

Inelastic neutron scattering experiments were performed on two inelastic chopper spectrometers: PHAROS at the Manuel Lujan Neutron Science Center (Los Alamos National Laboratory) and LRMECS at the Intense Pulsed Neutron Source (Argonne National Laboratory). The experimental configuration of LRMECS is the same as described previously.[29] The experimental configuration of PHAROS is similar to LRMECS with the notable exceptions of position sensitive detectors which cover a larger angular range (-10° - 140°) and a much larger sample moderator distance (18 m) enabling higher resolution experiments. Experiments were performed at a variety of incident energies and temperatures to fully explore the magnetic contribution to the INS spectrum. We have taken advantage of the nondispersive nature of CEF excitations to sum up the low angle range of detectors to improve the statistics of the data.

There are a number of background subtraction methods that may be used in order to extract the magnetic contribution to the INS spectra. Here we describe two of the most frequently used. Method one relies on subtracting the scattering observed in the nonmagnetic analog from that of the specified magnetic material as follows: $S_{mag}$ = S(Ce, SQ) – f S(NM,SQ) where SQ means small Q or low angle, NM means the nonmagnetic analog, and f is the ratio of the total scattering cross-section ($\sigma$) of the magnetic and nonmagnetic analog $\sigma(Ce)/\sigma(NM)$. For method two, the nonmagnetic analog is used to determine the scaling factor R = S(NM,LQ)/S(NM,SQ) between the high and low angle data (LQ represents large Q or high angle data). This same factor is then used to scale the high angle data (where nonmagnetic scattering dominates) to small angles (where magnetic scattering dominates) in the Ce compound. In the results reported here, method one has been used. We will discuss the effect of different background subtractions further in section IV.

To determine the CEF scheme, we have adopted two approaches. The first is to fit the magnetic contribution to the scattering to a CEF model for $Ce^{3+}$ in a tetragonal environment. Several datasets for different incident energies ($E_i$) and/or temperatures are fit simultaneously. The fitting parameters are: the CEF parameters ($B_l^m$'s), the width $\Gamma_{ie}$ of the inelastic excitations (which are modeled as Lorentzians), and a scale factor for each data set. We were unable to resolve a quasielastic contribution to the INS spectra and to prevent proliferation of fitting parameters we constrained the quasielastic width ($\Gamma_{qe}$) to be 1/4 of the inelastic width $\Gamma_{ie}$. For CeRhIn$_5$ (ref. 30) and CeCoIn$_5$ (ref. 28) this gives values of $\Gamma_{qe}$ that are in good agreement with estimates from NMR experiments.[31] (A variant on the nonmagnetic background subtraction methods is to allow the factors f or R to be variable parameters in the fits to the CEF model.)

We have used the parameters derived from these fits to calculate the magnetic susceptibility $\chi_{CEF}$. In these calculations, the CEF levels are treated as delta functions in energy. To account for the contribution of the 4f/conduction hybridization which is the source of the broadening of the CEF levels, we have added a mean field parameter $\lambda$ such that $1/\chi_{tot} = (1/\chi_{CEF}) + \lambda$. At high temperatures this is related to the Kondo temperature via $\lambda = T_K/C_{5/2}$ where $C_{5/2}$ is the Ce J = 5/2 Curie constant.

As an alternate method to account simultaneously for both the CEF and Kondo contributions to the INS spectra and to the susceptibility, we have carried out calculations for the Anderson impurity model for a $J = 5/2$ impurity in the presence of CEF using the non-crossing approximation. As in Ref. 30, we have used a Gaussian background band with halfwidth (at half maximum) 2.5 eV, setting the 4f level 2 eV below the Fermi level and including a spin orbit splitting 0.273 eV of the $J = 7/2$ states. The Kondo physics renormalizes the input CEF energies upwards by an amount approximately equal to $k_B T_K$ so the bare energies are chosen correspondingly smaller than those obtained from the CEF fits outlined above. The mixing parameter $\beta$ (eq. 2) and the 4f/conduction electron hybridization parameter V are then chosen to give reasonable fits to both the INS spectra and to the measured susceptibility.

## III. Results and Analysis

We now present the results of INS on CeMIn$_5$. We have made preliminary reports of some of these results elsewhere.[28,30,32] For the M = Co, Ir compounds we first present data that has been minimally processed and which conveys the presence of magnetic scattering in the INS spectra, but also serves as an indication of the uncertainty present in the measurements. (For CeRhIn$_5$ similar data has been reported elsewhere.[30]) We then present the magnetic portion of the scattering as well as the results of least squares fits to the CEF model and representative NCA calculations.

## A. CeCoIn$_5$

In fig. 1(a) INS spectra collected on PHAROS for CeCoIn$_5$ is contrasted to that of LaIrIn$_5$ with $E_i$ = 30.2 meV at 18 K. The spectra have been corrected for monitor counts, sample mass and neutron absorption, and the contribution of the empty sample holder has been removed. Extra intensity is evident in the INS spectra for CeCoIn$_5$ relative to the nonmagnetic analog LaIrIn$_5$. The extra intensity is attributed to CEF excitations in CeCoIn$_5$. Further evidence of CEF excitations in CeCoIn$_5$ is provided in fig. 1(b). Here we use data that has *not* been corrected for neutron absorption or for the sample holder scattering. We subtract the data for $E_i$ = 48.5 meV at 80 K from that taken at 10 K with each spectra normalized by the factor $n(\omega)+1 = \left(1-\exp\left(-\frac{\hbar\omega}{k_B T}\right)\right)^{-1}$ to account for the phonon population change with temperature. (This normalization only significantly affects the results at low energy transfers.) The fact that the difference shown in fig. 1(b) is positive on the energy loss side of the spectrum is characteristic of the presence of CEF excitations: as the occupation of the ground state doublet decreases with increasing temperature, the amplitude of the excitation from the ground state to the excited states also decreases. We conclude that two broad CEF excitations are present in the neutron scattering spectra for CeCoIn$_5$ centered at approximately 9 meV and 25 meV.

The magnetic part of the scattering, $S_{mag}$ (method 1) is displayed for CeCoIn$_5$ in fig. 2. The dependence of the magnetic formfactor has been removed so that the spectra represents the Q = 0 scattering. In fig. 2(a) the open circles and triangles indicate $E_i$ = 35 and 60 meV respectively at 10 K for INS spectra collected with LRMECS. Note that there are two broad peaks in $S_{mag}$ which is consistent with the previous assessment of the data presented in fig. 1. Fig 2(b) displays $S_{mag}$ for $E_i$ = 30.2 meV collected at 18 K on PHAROS. In both (a) and (b) the solid line represents a simultaneous fit to the CEF model for all three datasets. The resulting CEF parameters and other pertinent parameters are summarized in Table I. These results indicate that $\Gamma_7^1$ is the ground state, $\Gamma_7^2$ is the first excited state, and the second excited state is $\Gamma_6$. Due to the $\Delta J_z = \pm 1$ selection rule, the intensity of the peaks is sensitive to the degree of admixture of the $J_z = 5/2$ and $3/2$ states in the $\Gamma_7^1$ and $\Gamma_7^2$ states; in particular the strength of the 25 meV excitation ($\Gamma_7^1 \to \Gamma_6$) is proportional to $\beta$. The large widths of the inelastic excitations indicate the importance of strong Kondo spin fluctuations. In Fig. 3(a) we compare the measured magnetic susceptibility to the calculated value based on the CEF parameters determined from the INS data; the value of the mean field parameter $\lambda$ that accounts for the reduction of the susceptibility at high temperature due to the Kondo effect is given in Table I. In Fig. 2(a) and 3(a) we also present the results of the NCA Anderson impurity calculation with input parameters as given in Table I.

**B. CeIrIn$_5$**

Figure 4 displays similar INS spectra collected with PHAROS for CeIrIn$_5$ and LaIrIn$_5$. In fig. 4(a) the INS spectrum for CeIrIn$_5$ is contrasted to that of LaIrIn$_5$ with $E_i$ = 30.2 meV at 18 K. The spectra have been corrected for monitor counts, sample mass and neutron absorption, and the contribution of the empty sample holder has been removed. Additional intensity is observed for CeIrIn$_5$ relative to LaIrIn$_5$ which is attributed to CEF excitations. As in CeCoIn$_5$ more detail is provided upon examination of the temperature dependence of the scattering in CeIrIn$_5$; Fig 4(b) displays data taken with E$_i$ = 30.2 meV and 80 meV at 70 K subtracted from data taken at 18 K and 10K with the same normalization as in fig 1(b). Taken together figs. 4(a) and (b) indicate a broad CEF excitation centered in the 3-5 meV range and a second excitation near 30 meV.

The magnetic contribution to the INS spectra of CeIrIn$_5$ is displayed in fig. 5. For clarity, the data has been smoothed. (Unsmoothed data as well as the magnetic part of the PHAROS INS spectra are reported in Ref. 32.) The symbols in fig. 5 for three different E$_i$'s indicate two very broad peaks in the INS which are attributed to CEF excitations. The solid line in fig. 5 is a fit to a CEF model similar to the one presented for CeCoIn$_5$ above. The resulting parameters of this fit are summarized in table I. As in the case of CeCoIn$_5$, $\Gamma_7^1$ is the ground state, $\Gamma_7^2$ is the first excited state, and the second excited state is $\Gamma_6$. The CEF parameters also reproduce the high temperature magnetic susceptibility as shown in fig. 3(b). The CEF splitting (6.7 meV) is somewhat smaller for the first excited state in CeIrIn$_5$ than in CeCoIn$_5$, however $\Gamma_{ie}$ is somewhat larger: 8.7 meV as compared to 6.6 meV for CeCoIn$_5$. The results of NCA calculations are included in figs. 3(b) and 5 as dashed lines.

## C. CeRhIn$_5$

Figure 6(a) shows the results of subtracting the data for CeRhIn$_5$ at 80 K from that taken at 10 K; CEF excitations are apparent at 7 and 25 meV. Figure 6(b) shows S$_{mag}$ for CeRhIn$_5$. (We have obtained similar data with PHAROS which confirms the results of ref. 30.) The 25 meV peak intensity is smaller relative to the 7 meV peak than in either CeCoIn$_5$ or CeIrIn$_5$ indicating that CeRhIn$_5$ must have the least admixture of the $J_z = 3/2$ state in the $\Gamma_7^1$ ground state. Note that the admixture of $J_z = 3/2$ is not zero, since the peak intensity of the 25 meV excitation then would be identically zero, which is clearly not the case. The solid line in fig. 6(b) indicates the best fit to a CEF model with parameters as summarized in table I. The relative ordering of the wavefunctions is the same for CeRhIn$_5$ as for CeCoIn$_5$ and CeIrIn$_5$. The results of the NCA calculations are shown in Fig. 3(c) and 6(b).

## IV. Discussion

We first discuss the systematic errors in our determination of the CEF parameters. Since absorption is strong in these compounds, the signals are weak; under these circumstances it is possible to overestimate the linewidth of broad peaks. Since the excitations in CeRhIn$_5$ and CeCoIn$_5$ are reasonably well-resolved, we don't think this is a problem, and hence we argue that the large observed linewidths, especially in CeIrIn$_5$, are not artifacts of the analysis but are real effects. Absorption also affects the estimate of the strength of the $\Gamma_7^1 \rightarrow \Gamma_6$ transition, since the final neutron energies are small for these larger energy transfers. Our absorption correction is based on a flat plate sample geometry and errors could arise from variation in sample thickness. This leads to unknown uncertainty in our estimate of β. The determination of the nonmagnetic scattering also leads to systematic uncertainty; the methods discussed above are reasonable but not rigorous. To estimate the resulting systematic uncertainty, we have examined the range of parameters obtained for all methods of nonmagnetic scattering subtraction described above (methods one and two and their variants where R and f are allowed to vary) and for various combinations of datasets (different combinations of $E_i$ and T). We find for all compounds that variations in β are small (±0.05) and variations in $E[\Gamma_7^1 \rightarrow \Gamma_6]$ are of order ±2 meV. Variations in $E[\Gamma_7^1 \rightarrow \Gamma_7^2]$ are small (±0.2 meV) for CeRhIn$_5$, larger (±1 meV) for CeCoIn$_5$ and largest for CeIrIn$_5$ where, due to the large inelastic linewidth, the excitations are not well resolved so that values of $E[\Gamma_7^1 \rightarrow \Gamma_7^2]$ in the range 2-7 meV all give reasonable fits to the INS spectra and susceptibility. In all cases, the estimates of systematic error are larger than the statistical error. Consequently, error bars are not given in table I.

The values we report in table I can be viewed as representative, within these limits. They are obtained on the same spectrometers, under identical conditions, with identical methods of absorption correction and nonmagnetic background subtraction (method one). Hence the results are consistent between the compounds and should accurately reflect trends in the CEF parameters. Our method of subtracting high temperature from low temperature raw data (Figs. 1(b), 4(b) and 6(a)) confirms that the positions of the peaks given in Table I are essentially correct and, since the ratio of the $\Gamma_7^1 \rightarrow \Gamma_6$ peak intensity to that of the $\Gamma_7^1 \rightarrow \Gamma_7^2$ peak increases in the sequence M = Rh, Ir, Co, it also confirms the trend seen in Table I that β increases in the same sequence. In addition, the calculations of the susceptibility based on the parameters of Table I adequately represent the magnitude and anisotropy of the susceptibility for T > 50-100K. These calculations include a single mean-field parameter λ which accounts for the reduction of the susceptibility by the Kondo effect (and also by antiferromagnetic correlations) at high temperatures. The values of $T_K$ obtained from the assumption $\lambda = T_K/C_{5/2}$ are 28, 32 and

56 K for M = Rh, Co, Ir respectively. These can be viewed as high temperature Kondo temperatures, in the regime where the excited states are occupied. As such, they are not only reasonable, but they also show the same trend as the inelastic linewidths, to which they should be proportional. We emphasize here that such CEF calculations of the susceptibility, which treat the CEF levels as delta functions in energy and include the hybridization through λ can only be valid at elevated temperatures and cannot be expected to capture the low temperature Kondo physics and the low temperature magnetic correlations which are important in these compounds. Given all this, we believe that the values of excitation energies, mixing parameters and linewidths given in table I are essentially correct, within the limits of systematic error discussed above.

Moreover, our results are consistent with a picture where the CEF parameters of $CeMIn_5$ are derived from the parent compound $CeIn_3$. In $CeIn_3$, the ground state is found to be a $\Gamma_7$ doublet while the excited state at ~12 meV is found to be a $\Gamma_8$ quartet.[24][25][26] Upon lowering the cubic symmetry of $CeIn_3$ to the tetragonal symmetry of $CeMIn_5$ the 4-fold degeneracy of the $\Gamma_8$ quartet is lifted, resulting in a CEF level scheme of three doublets. The ground state remains the $\Gamma_7$ (denoted as $\Gamma_7^1$ in tetragonal symmetry), the first excited state ($\Gamma_7^2$) and the upper CEF doublet ($\Gamma_6$) are derived from the $\Gamma_8$ quartet. For $CeCoIn_5$ the mixing parameter β remains close to the value $\sqrt{\frac{5}{6}}$ that it has for cubic symmetry, but for the other two compounds, β is affected by the tetragonal CEF. In addition, $\Gamma_{ie}$ for $CeIn_3$ (3 meV[26]) is similar to the value of 2.3 meV found for $CeRhIn_5$, while values of 6.6 and 8.7 meV are found for $CeCoIn_5$ and $CeIrIn_5$ respectively.

The results of all previous attempts to elucidate the CEF parameters in CeMIn$_5$ are summarized in table II. The older results are based on using susceptibility χ, specific heat C$_v$ and thermal expansion α and are relatively insensitive to the upper excitation $\Delta_2$ because the CEF contributions to χ, C$_v$ and α are small at the higher temperatures where this excitation becomes thermally populated. We first discuss the three previous attempts at identifying the CEF level scheme in CeCoIn$_5$. All identify a $\Gamma_7^1$ ground state except for the case of Ref. 17 and 19. These latter authors identify a $\Gamma_6$ ground state and a positive value for $B_2^0$. In tetragonal symmetry at high temperatures in the absence of the Kondo effect or magnetic correlations, the parameter $B_2^0$ should be proportional to $(1/\chi_{ab})$ - $(1/\chi_c)^{33}$, and hence for these compounds should be negative. To account for this in their analysis of the magnetic susceptibility an anisotropic mean field parameter was included. The results of ref. 22 indicate a $\Gamma_7^1$ ground state; however, they find β = 0.519 indicating a larger admixture of the J$_z$ = 5/2 state into the ground state than the determination here; they find somewhat different energy splittings as well. Of the CEF schemes for CeCoIn$_5$ in the literature the results of ref. 21 show a similar β to that determined here, but some difference in the actual value of the splittings.

There have been fewer attempts to identify the CEF level scheme in CeIrIn$_5$. Here the disagreement is substantially less than in CeCoIn$_5$. The value of β found by ref. 20 is significantly different from the value of the present work, though the energy level splittings are in reasonable agreement. Although ref. 17 finds a different groundstate, we note that only the modulus of the matrix elements is observable in experimental probes of CEF excitations and consequently a distinction cannot be made between the $\Gamma_7^1$ and $\Gamma_7^2$ wavefunctions with equal weights of $J_z$ = 5/2 and 3/2 (this corresponds to changing the sign of $B_4^4$). Thus the ground state of ref. 17 is similar to the one found in this paper using their value of β. However, they find the first excited state to be a $\Gamma_6$ and they find a much smaller groundstate upper level splitting.

There have been somewhat fewer attempts to determine the CEF level scheme in CeRhIn$_5$. Of these, all agree on the relative ordering of the CEF levels. However, both refs. 17 and 20 find small values of β. As mentioned previously, this would indicate an even smaller intensity for the upper excitation in the INS spectra and therefore such small values of β can be discarded. Furthermore, calculations based on these CEF schemes give an in-plane magnetic moment $g\mu_B \langle J_x \rangle$ which is a factor of two smaller than the in-plane ordered moment observed by neutron diffraction (0.75).[34] This cannot be the case indicating a larger admixture of 5/2 and 3/2 consistent with the moment of 0.92 $\mu_B$ calculated from our results and indicating a degree of moment reduction due to the Kondo effect.

We now turn to a discussion of the NCA calculations. It is clear from the broadness of the CEF excitations in CeMIn$_5$ that Kondo spin fluctuations play an essential role. Both the fits to the INS spectra with a CEF model plus Lorentizian widths and the fits to the magnetic susceptibility with a CEF contribution and the addition of a mean field constant only represent the effect of Kondo spin fluctuations in an *ad hoc* way. In principle the NCA calculations capture the Kondo physics, but not the effect of magnetic correlations and/or 4f lattice coherence. The values of V = 469, 470, 456 meV for M = Co, Ir, Rh respectively (table I) indicate that a rather modest increase in hybridization leads to the more significant changes in linewidths of the CEF excitations. We note that the NCA calculations are unable to reproduce the low temperature features in the magnetic susceptibility. For this reason, and for the reason that we can find no set of CEF parameters which reproduces both the neutron data and the plateau in the c-axis magnetic susceptibility in $CeCoIn_5$, we agree with the conclusions of ref. 21 that the feature must be due to correlations.

From table I it can be seen that there is no correlation between the magnitude of $\Delta_1$ and the superconducting transition temperature, as suggested by recent theory.[18] However, as the superconducting transition temperature increases so does the mixing parameter β. Moreover, we find that a significant amount of hybridization appears to be required for the formation of the superconducting state as can be seen from the values of $\Gamma_{ie}$ in table I. A similar behavior is observed on comparison of $CeCu_2Ge_2$ (antiferromagnetic at 4.1 K[35]) and $CeCu_2Si_2$ (superconducting at 0.5 K[36]): $CeCu_2Si_2$ has a larger $\Gamma_{ie}$ than $CeCu_2Ge_2$ (refs. 35 and 37) and tuning the hybridization of $CeCu_2Ge_2$ with applied pressure results in a superconducting transition of 0.64 at 101 kbar.[38] This also is consistent with the behavior of $CeIn_3$ under application of pressure[16] and as mentioned above, $\Gamma_{ie}$ for the CEF excitation in antiferromagnetic $CeIn_3$ is consistent with the value found in antiferromagnetic $CeRhIn_5$ rather than the much large values of $\Gamma_{ie}$ found in either of the ambient pressure superconductors $CeCoIn_5$ or $CeIrIn_5$. With the application of pressure, the hybridization is tuned suppressing the antiferromagnetic order. The superconducting state is then formed near the QCP where antiferromagnetic order is suppressed. The evolution of the hybridization in CeMIn$_5$ indicates a similar picture where the substitution of a different transition metal is sufficient to change the hybridization. This suggests that CeCoIn$_5$, for which the superconducting transition temperature is highest and $\Gamma_{ie}$ is fairly large, is near the QCP, while CeRhIn$_5$, which at ambient pressure is magnetically ordered and for which $\Gamma_{ie}$ is relatively small, is on the magnetic side of the QCP. CeIrIn$_5$, where the superconducting transition temperature is lower and $\Gamma_{ie}$ is larger than in CeCoIn$_5$, is slightly farther out on the nonmagnetic side of the QCP phase diagram. The QCP picture has been advocated by a number of previous authors.[11][12][13][14][15]

## V. Conclusions

We have measured the CEF excitations of the CeMIn$_5$ (M = Co, Rh, Ir) series of heavy fermion superconductors by means of INS. The CEF excitations are broadened by the effect of Kondo spin fluctuations. Consequently, we have adopted two approaches to determine the CEF parameters, energy level splittings, and wavefunctions. The first approach fits the magnetic portion of the INS spectra by a CEF model where the peak widths are represented by a Lorentzian line shape. The second approach utilizes NCA calculations and represents a more sophisticated means of accounting for the effect of Kondo spin fluctuations. Both of these methods are able to reproduce the INS data and the magnetic susceptibility and the resulting CEF parameters, level splittings and wavefunctions have been tabulated and compared to the results of previous workers. Furthermore, these approaches yield a picture in which the CEF level splitting in all three materials is similar and can be thought of as being derived from the cubic parent compound CeIn$_3$ in which an excited state quartet at ~12 meV is split into two doublets by the lower symmetry of the tetragonal environment of CeMIn$_5$. Although we find no correlation between the superconducting transition temperature and the level splitting, we do find a correlation between the f conduction electron hybridization and the superconducting transition temperature where significant hybridization is required for the formation of the superconducting state.

Acknowledgments: We acknowledge fruitful discussions with W. Bao and S. Kern. Work at Irvine was supported by the Department of Energy (DOE) under Grant No. DE-FG03-03ER46036. Work at Temple U. was supported by the DOE under Grant No. DE-FG02-01ER45827. Work at Los Alamos and Argonne was performed under the auspices of the DOE.

# Tables

Table I. CEF Parameters for CeMIn$_5$. Except for $\chi^2$ and $\beta$ which are dimensionless and $\lambda$ which is given in (mol/emu) all units are meV. In all three materials the groundstate is a $\Gamma_7^1$, the first excited state a $\Gamma_7^2$, and the second excited state is a $\Gamma_6$. The numbers in square brackets are from the NCA calculations.

|  | CeRhIn$_5$ | CeCoIn$_5$ | CeIrIn$_5$ |
|---|---|---|---|
| $\chi^2$ | 0.69 | 0.52 | 0.83 |
| $B_2^0$ | -1.03 | -.81 | -1.2 |
| $B_4^0$ | 0.044 | 0.058 | 0.06 |
| $B_4^4$ | 0.122 | 0.139 | 0.12 |
| $\beta$ [NCA] | 0.60 [0.6] | 0.86 [0.95] | 0.70 [0.71] |
| $\Gamma_{ie}$ | 2.3 | 6.6(4) | 8.7 |
| $E(\Gamma_7^2)$ [NCA] | 6.9 [7] | 8.6 [6.45] | 6.7 [2] |
| $E(\Gamma_6)$ [NCA] | 24 [25] | 25 [21.44] | 29 [22.56] |
| V | 456 | 469 | 470 |
| $\lambda$ | 35 | 40 | 70 |

**Table II.** CEF Level Schemes (comparison of all determinations). $\Delta_1$ and $\Delta_2$ represent the energy splitting between the ground state and first and second excited states respectively. The column corresponding to Order indicates the order of wavefunctions from the groundstate to the upper level. The column corresponding to β indicates the value for the groundstate wavefunction. Because of inconsistency in the literature with respect to the labeling of the wavefucntions we have defined the wavefunctions consistent with eq. 2.

| Ref. | $\Delta_1$ (meV) | $\Delta_2$ (meV) | Order | β | Method |
|---|---|---|---|---|---|
| CeCoIn$_5$ | | | | | |
| 17 and 19 | 2.8 | 8.8 | $\Gamma_6\ \Gamma_7^1\ \Gamma_7^2$ | 0.1 | Mag. Susc., Spec. Heat, and NMR |
| 21 | 13 | 17 | $\Gamma_7^1\ \Gamma_7^2\ \Gamma_6$ | 0.92 | Mag. Susc. and Spec. Heat |
| 22 | 13 | 14.3 | $\Gamma_7^1\ \Gamma_7^2\ \Gamma_6$ | 0.519 | Mag. Susc. |
| present | 8.6 | 25 | $\Gamma_7^1\ \Gamma_7^2\ \Gamma_6$ | 0.86 | INS |
| CeIrIn$_5$ | | | | | |
| 17 | 3.9 | 10.8 | $\Gamma_7^1\ \Gamma_6\ \Gamma_7^2$ | 0.66 | Mag. Susc. and Spec. Heat |
| 20 | 5.3 | 25.9 | $\Gamma_7^1\ \Gamma_7^2\ \Gamma_6$ | 0.213 | Mag. Susc. and Therm. Expan. |
| present | 6.7 | 29 | $\Gamma_7^1\ \Gamma_7^2\ \Gamma_6$ | 0.70 | INS |
| CeRhIn$_5$ | | | | | |
| 17 | 6.0 | 12.1 | $\Gamma_7^1\ \Gamma_7^2\ \Gamma_6$ | $\approx 0$ | Mag. Susc. and Spec. Heat |
| 20 | 5.9 | 28.5 | $\Gamma_7^1\ \Gamma_7^2\ \Gamma_6$ | 0.247 | Mag. Susc. and Therm. Expan. |
| present | 6.9 | 24 | $\Gamma_7^1\ \Gamma_7^2\ \Gamma_6$ | 0.60 | INS |

**Figure Captions**

Fig 1. (a) Inelastic Neutron Scattering Spectra for $CeCoIn_5$ and $LaIrIn_5$. Solid squares (open circles) indicate the spectrum for $CeCoIn_5$ ($LaIrIn_5$) collected at 18 K with $E_i = 30.2$ meV. (b) Temperature dependence of the INS response of $CeCoIn_5$. Shown is data collected at 80 K subtracted from data collected at 18 K with $E_i = 48.5$ meV where the spectra at each temperature is normalized as described in the text.

Fig. 2. $S_{mag}$ for $CeCoIn_5$. (a) The open circles (triangles) indicate $S_{mag}$ for data collected on LRMECS with $E_i = 35$ and 60 meV respectively. (b) Solid squares for data collected on PHAROS with $E_i = 30$ meV. In both (a) and (b) the solid line indicates a simultaneous fit to a CEF model. The dashed line in (a) is the result of an NCA calculation as described in the text.

Fig. 3. The magnetic Susceptibility for $CeMIn_5$. In all panels circles (triangles) represent $\chi_c$ ($\chi_{ab}$) and solid (dashed) lines represent a CEF model fit (NCA calculations) respectively. The parameters of the fits are given in Table I.

Fig. 4. (a) Inelastic Neutron Scattering Spectra for $CeIrIn_5$ and $LaIrIn_5$. Solid squares (open circles) indicate the spectrum for $CeIrIn_5$ ($LaIrIn_5$). (b) Temperature dependence of the INS response of $CeIrIn_5$. Shown is data collected at 80 K subtracted from data collected at 18 K with $E_i = 30.2$ meV on PHAROS and data collected at 70 K subtracted from data collected at 10 K with $E_i = 80$ meV on LRMECS where the spectra at each temperature is normalized as described in the text.

Fig. 5. $S_{mag}$ for $CeIrIn_5$. The open circles, closed circles, and open diamonds indicate $S_{mag}$ for data collected at 10 K on LRMECS with $E_i = 15, 30$ and 60 meV respectively. The solid line indicates a simultaneous fit to a CEF model. The dashed line is the result of an NCA calculation as described in the text.

Fig. 6. (a) Temperature dependence of the INS response of $CeRhIn_5$. Shown is data collected at 100 K subtracted from data collected at 10 K with $E_i = 35$ meV on LRMECS where the spectra at each temperature is normalized as described in the text. (b) The solid line indicates a simultaneous fit to a CEF model with $E_i = 35$ meV at temperatures of 10, 70, and 140 K and $E_i = 60$ meV at 10 K as described in ref. 30. The dashed line in (b) is the result of an NCA calculation as described in the text.

**Figures**

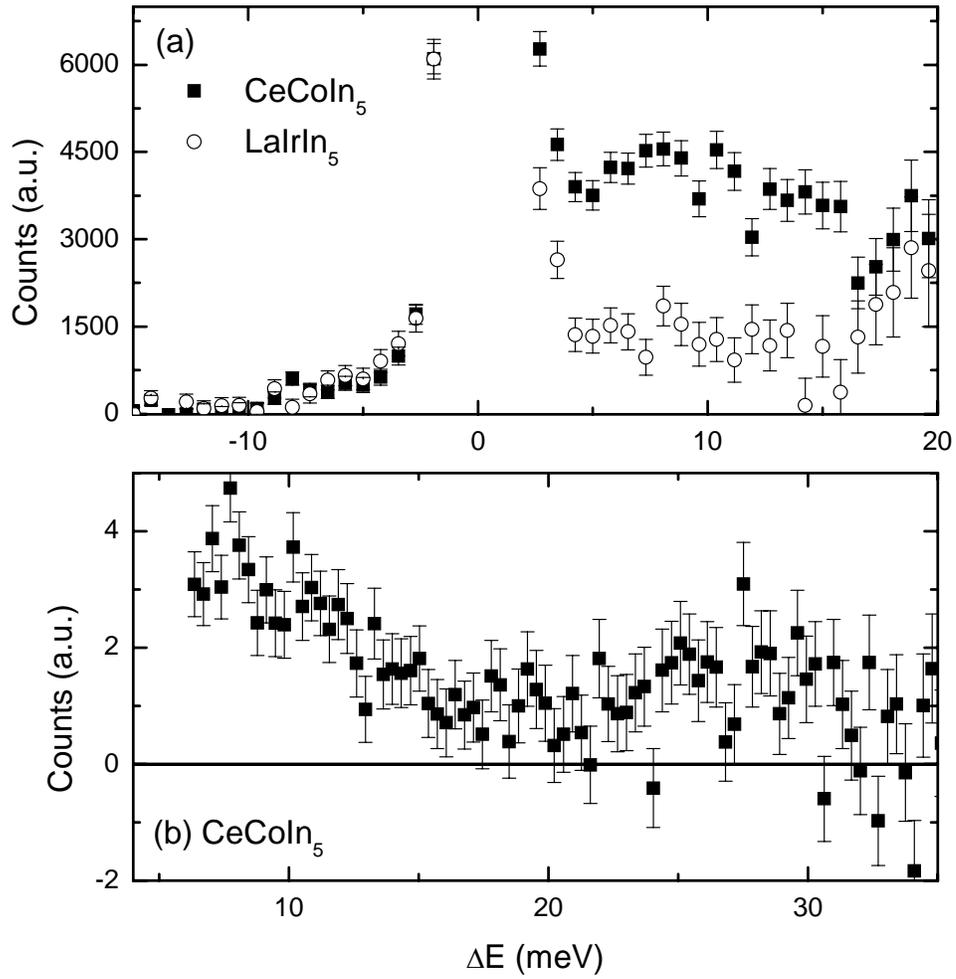

Figure 1 A.D. Christianson *et al*.

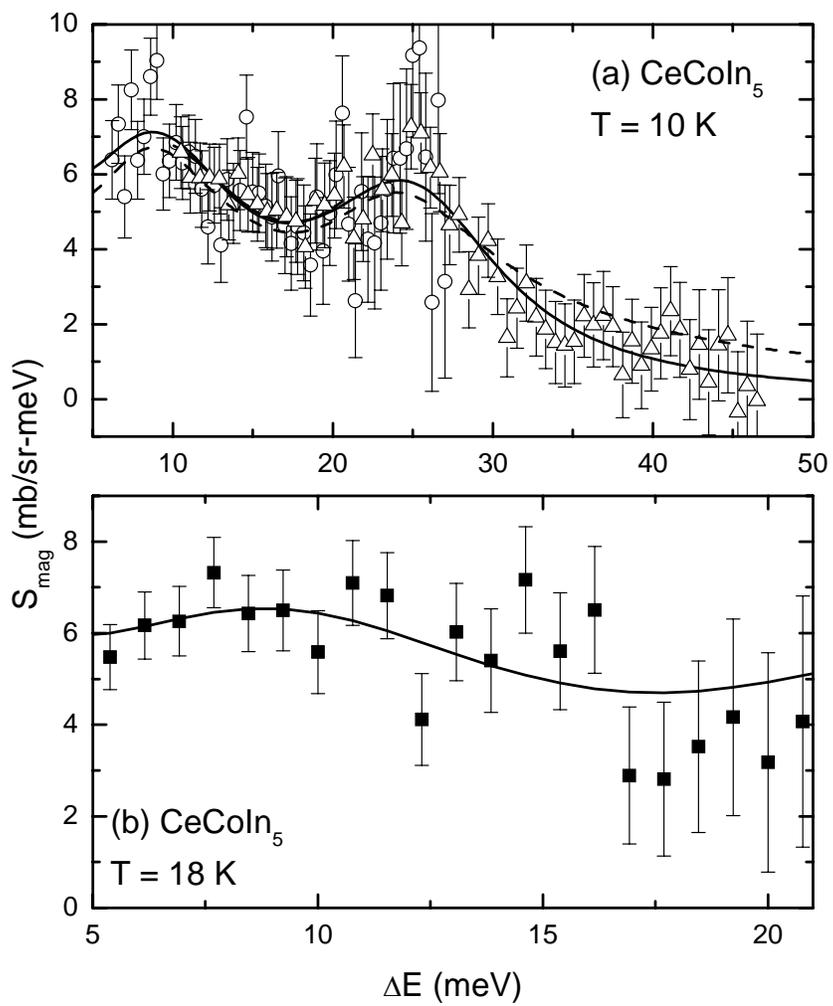

Figure 2 A. D. Christianson *et al*.

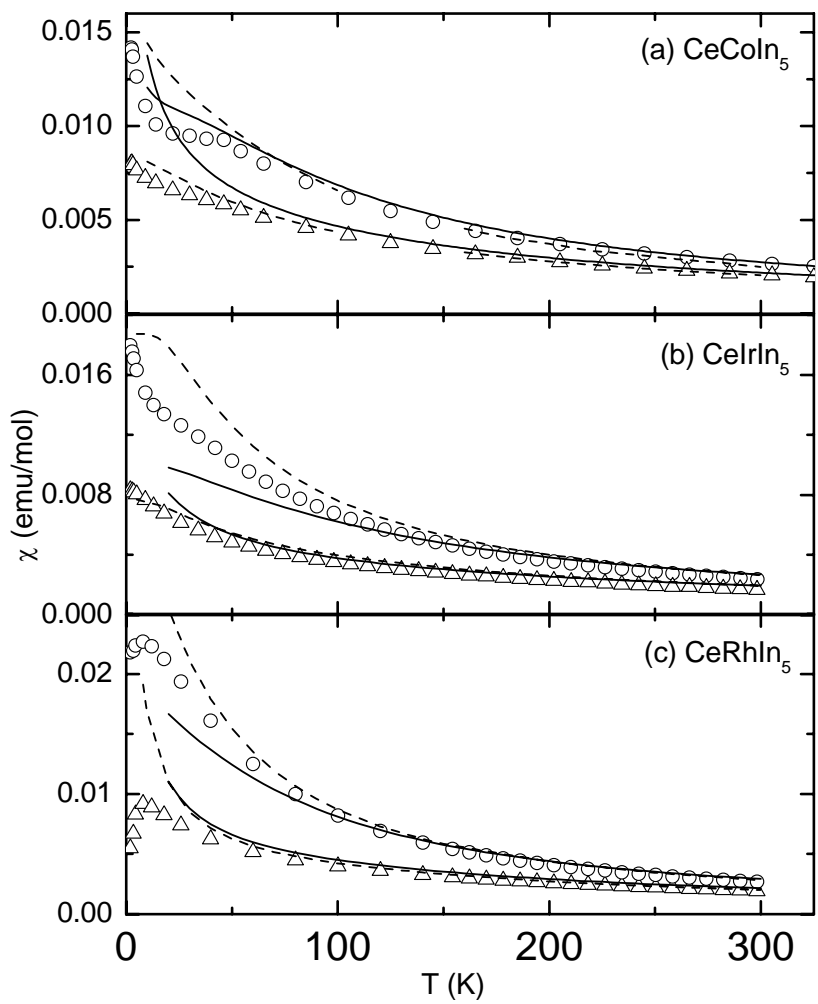

Figure 3 A. D. Christianson *et al*.

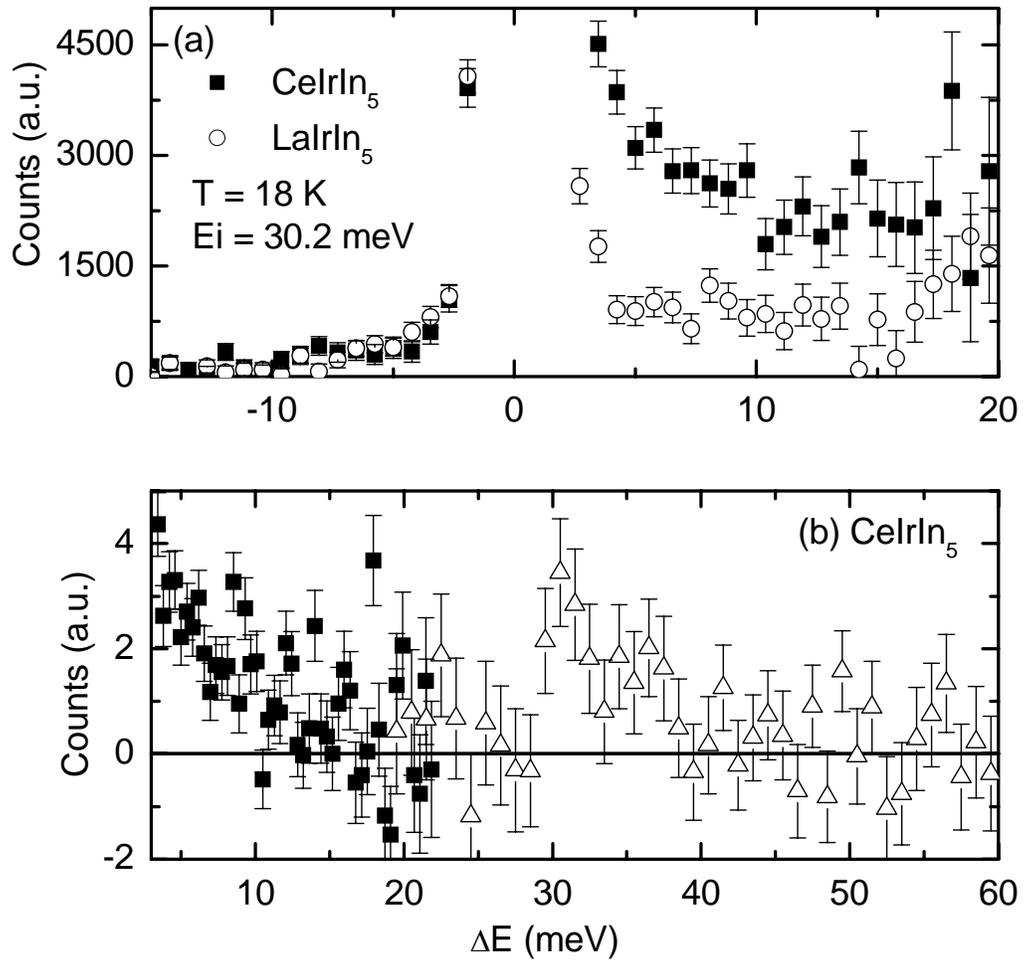

Figure 4 A. D. Christianson *et al*.

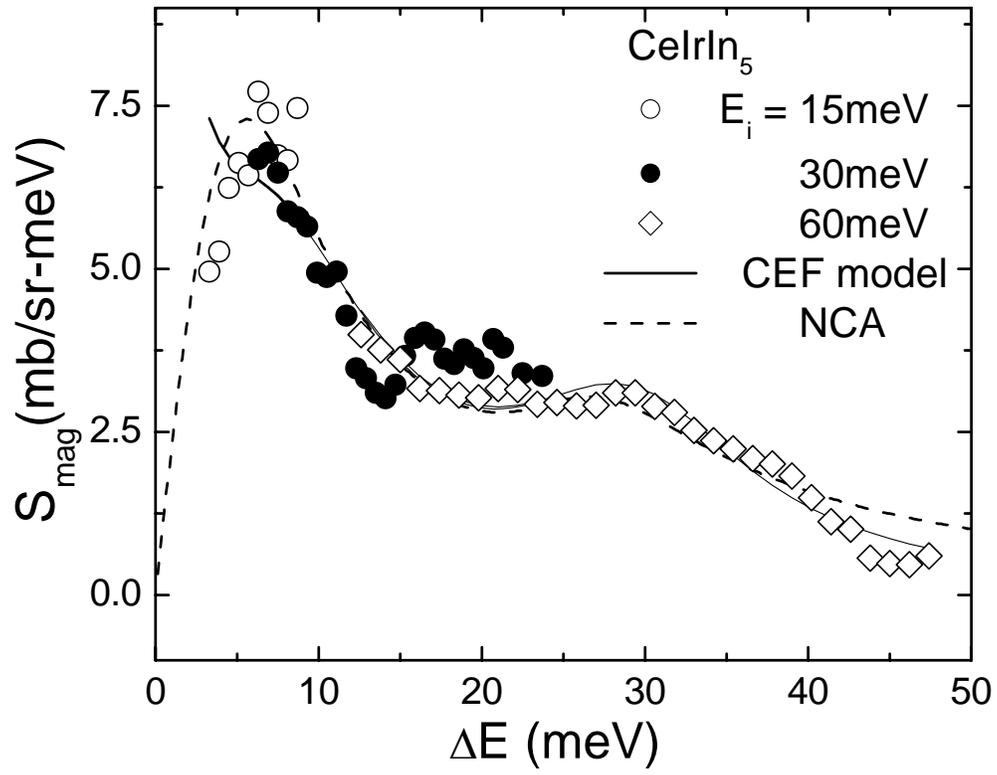

Figure 5 A. D. Christianson *et al*.

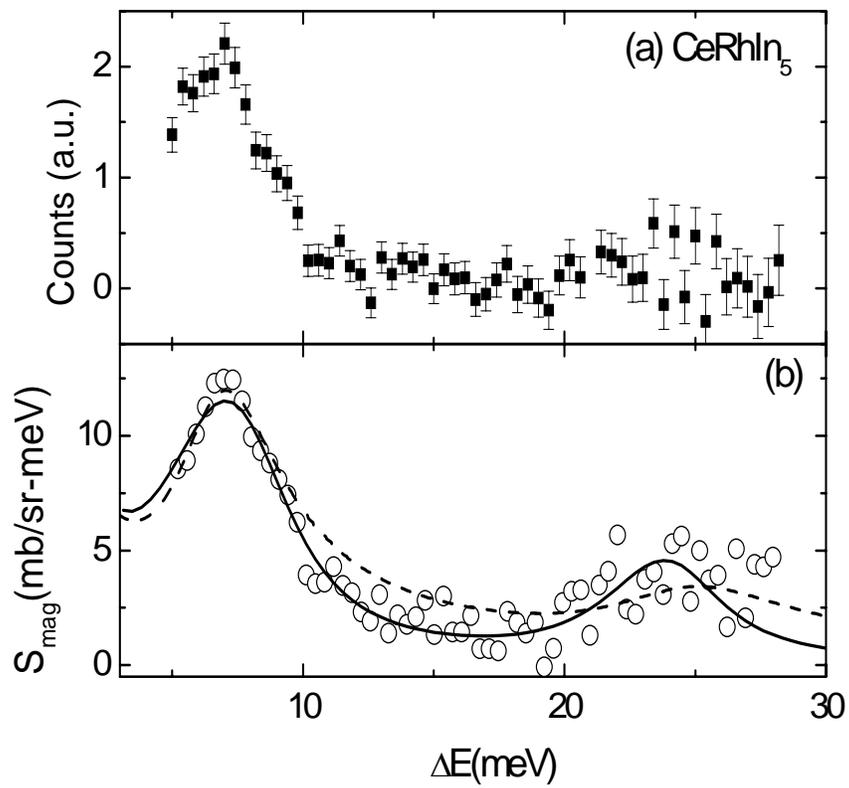

Figure 6 A. D. Christianson *et al*.